\documentclass[reprint,aps,prl,showpacs,
superscriptaddress,longbibliography]{revtex4-1}
\usepackage{graphicx}
\usepackage{amsmath}
\usepackage{amssymb}
\usepackage{epstopdf}
\usepackage{amsfonts}
\usepackage{dcolumn}
\usepackage{bm}
\usepackage{color}

\begin{document}

\title{Cluster formation in two-component Fermi gases}
\author{X. Y. Yin}
\affiliation{Centre for Quantum and Optical Science, Swinburne University of Technology,
Melbourne, Victoria 3122, Australia}
\author{Hui Hu}
\affiliation{Centre for Quantum and Optical Science, Swinburne University of Technology,
Melbourne, Victoria 3122, Australia}
\author{Xia-Ji Liu}
\affiliation{Centre for Quantum and Optical Science, Swinburne University of Technology,
Melbourne, Victoria 3122, Australia}

\date{\today }

\begin{abstract}
Two-component fermions are known to behave like a gas of molecules in the limit of Bose-Einstein condensation 
of diatomic pairs tightly bound with zero-range interactions. 
We discover that the formation of cluster states occurs when the effective range of two-body
interaction exceeds roughly $0.46$ times the scattering length,
regardless of the details of the short-range interaction.
Using explicitly correlated Gaussian basis set expansion approach, 
we calculate the binding energy of cluster states in trapped few-body systems
and show the difference of structural properties between cluster states and
gas-like states. We identify the condition for cluster formation
and discuss potential observation of cluster states in experiments.
\end{abstract}

\maketitle

{\em Introduction:} 
A fermion is a particle that follows Fermi-Dirac statistics, which gives rise to Pauli exclusion principle~\cite{fermi34}.
The Pauli exclusion principle states that two or more identical fermions cannot occupy the same quantum state
within a quantum system. The result is the emergence of Fermi pressure that prevents white dwarfs and neutron stars
from gravitational collapse~\cite{whitedrawf}. 

Fermi pressure is also responsible for stabilizing dilute two-component Fermi gases~\cite{petrov04, giogini04, baym06, blumerpp}. 
As the strength of two-body interaction changes, the two-component Fermi gases experience a
crossover from weakly correlated Bardeen-Cooper-Schrieffer (BCS) pairing to a Bose-Einstein condensate (BEC) of 
tightly bound pairs~\cite{jin04, ketterle04}.
In the BEC limit of the BCS-BEC crossover, two unlike fermions form a molecule, which is a composite boson~\cite{giogini04, petrov04}.
The system is then governed by dimer-dimer interactions between molecules.
Structureless bosons with two-body interactions are known to form cluster states, such as three-body Efimov states~\cite{greenermp}. 
However, such cluster states have not yet been found in two-component Fermi gases in the BEC limit.
Previous numerical calculations in few-body systems~\cite{blume09, blume11} and larger systems~\cite{blume07, giogini04}
focused on the zero-range limit and no cluster states were found.
The reason is that the Fermi pressure between identical fermions prevents such cluster states to form.
However, cluster states could in principle exist with finite-range interactions.
In this Letter, we show that the formation of cluster states occurs when the effective range of the
two-body interaction exceeds roughly $0.46$ times the scattering length, regardless of the details of the short-range interaction.

The identification of such cluster states are crucial in three ways.
First, the formation of clusters in small two-component Fermi gases in three dimension (3D) corresponds to a 
phase transition from a droplet-like phase to a gas-like phase in many-body systems. 
This is reminiscent of the Luttinger liquid and gas of molecule transition in one dimension~\cite{law08}.
Second, the condition for cluster formation is important for preparing such systems in the lab.
The magnitude of effective range can approach that of the scattering length in the vicinity of a
Feshbach resonance~\cite{hulet13}.
Current technology allows preparing Fermi gases interacting through large effective range~\cite{feshbachrmp, ohara12}.
The stability of the system and the atom loss rate will be strongly affected by the cluster formation.
Third, two-component Fermi gases share similarities with the low-density regions of a neutron star interior
where the scaled interaction strength varies and the effective range is, in general, not negligible~\cite{neutronstar}.
Therefore, the parameter regime in certain parts of the neutron star can overlap with our study. Our results can
provide insights in stability of local regions inside a neutron star.

Previous studies discussed the instability of trapped fermionic gases with attractive interactions within the
mean-field frame work and the range of interaction was identified as an important factor~\cite{baym06}. 
Here, we pursue an $ab$ $initial$ calculation using explicitly correlated Gaussian (ECG) basis set expansion approach to
study small systems consisting up to 6 particles. 
The microscopic approach has been proven successful in understanding 
physics in larger systems~\cite{blumerpp, parish17}.

{\em System Hamiltonian:} 
We consider equal mass two-component Fermi gases in 3D consisting of $N_\uparrow$ spin-up and $N_\downarrow$ spin-down
atoms ($N_\uparrow=N_\downarrow=N$) under external spherically symmetric harmonic confinement with angular trapping frequency $\omega$.
The system Hamiltonian $H$ reads
\begin{eqnarray}
\label{eq_H}
H=\sum_{i_\uparrow=1}^{N} ( -\frac{\hbar^2}{2m}
\nabla_{i_\uparrow}^2
+ \frac{1}{2}m\omega^2 \vec{r}_{i_\uparrow}^2 )
+\sum_{i_\downarrow=1}^{N} (-\frac{\hbar^2}{2m} 
\nabla_{i_\downarrow}^2  \nonumber \\
+\frac{1}{2} m\omega^2 \vec{r}_{i_\downarrow}^2 )
+\sum_{i_\uparrow=1}^{N}\sum_{i_\downarrow=1}^{N}
V_{2b}(r_{i_\uparrow i_\downarrow}),
\end{eqnarray}
where $m$ denotes the mass of a single atom,
$\vec{r}_{i_\uparrow}$ and $\vec{r}_{i_\downarrow}$ denote
the position vector of the $i$th spin-up and down atom with respect to the
trap center, respectively.
We define the harmonic oscillator length $a_\text{ho}=\sqrt{\hbar/(m\omega)}$
and harmonic oscillator energy $E_\text{ho}=\hbar\omega$.
Interspecies two-body interaction
potential $V_{2b}$ depends on the interparticle distance 
$r_{i_\uparrow i_\downarrow}$, $r_{i_\uparrow i_\downarrow}=|\vec{r}_{i_\uparrow}-\vec{r}_{i_\downarrow}|$.

\begin{figure}[htbp]
\includegraphics[angle=0,width=85mm]{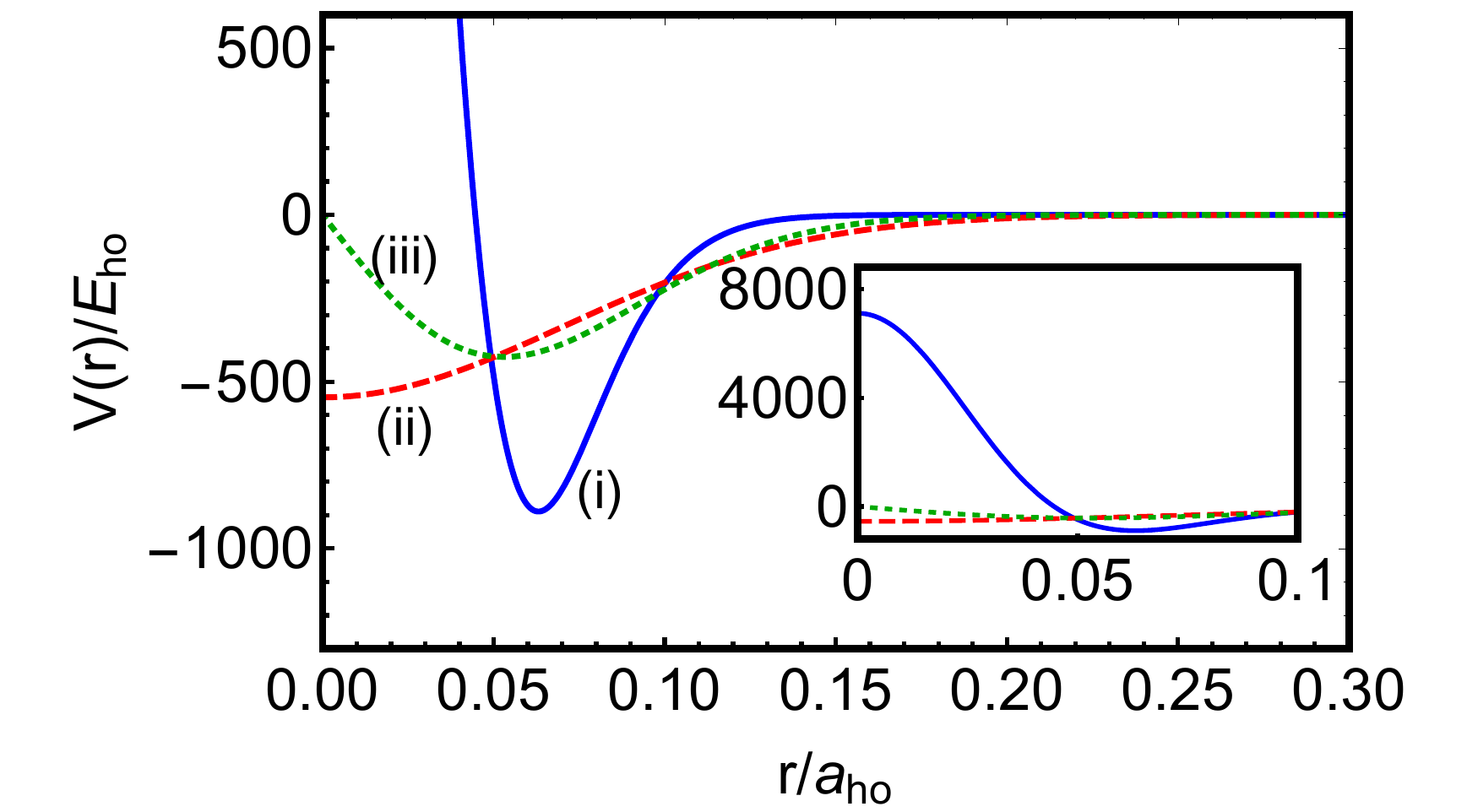}
\caption{(Color online) Solid, dashed, and dotted lines
show the short range potential curves (i), (ii) and (iii), respectively. The value of $a_s$ and $r_\text{eff}$
coincides with the condition for
cluster formation presented later in Fig.~\ref{fig_phase}. Inset shows the same plot on a different scale.
}
\label{fig_potential}
\end{figure}

We consider three different short-range potentials:
(i) an attractive Gaussian potential with a repulsive core,
$V_{2b}^{(i)}(r)=V_0 \exp (-r^2/4r_0^2)-2V_0 \exp (-r^2/2r_0^2)$;
(ii) an attractive Gaussian potential,
$V_{2b}^{(ii)}(r)=V_0 \exp (-r^2/2r_0^2)$;
and (iii) a modified attractive Gaussian potential,
$V_{2b}^{(iii)}(r)=rV_0 \exp (-r^2/2r_0^2)$.
The scattering phase shift of two particles at low energy can
be expanded as $-k\cot[\delta(k)]=1/a_s-r_\text{eff}k^2/2$, where $k$ is the wave vector,
$a_s$ the $s$-wave scattering length, and $r_\text{eff}$ the effective range.
For a fixed $r_0$, we adjust $V_0$ such that
$V_{2b}(r)$ has a certain $a_s$.
For all three types considered, shallow attractive potentials
do not support two-body $s$-wave bound state in free-space. As the depth increases,
the potential supports successively more two-body bound states~\cite{blumerpp}.
In this work, we only consider attractive potentials that support one $s$-wave bound state in free space
(see Supplemental Material~\cite{supp}).
We then calculate $r_\text{eff}$ for such potential.
Solid, dashed, and dotted lines in Fig.~\ref{fig_potential} show potentials (i), (ii), and (iii) that produce
the same $a_s=0.2a_\text{ho}$ and
$r_\text{eff}=0.09a_\text{ho}$, respectively.
We will show later that the condition for cluster formation does not strongly depend on
the type of potential. This indicates that the physics is well described by the two parameters
$a_s$ and $r_\text{eff}$ alone.
Although all three types are short-range,
potential (i) simulates the repulsive core in realistic atom-atom
interactions and can produce a larger range of $r_\text{eff}$ for a fixed $a_s$.

We solve the time-independent Schr\"odinger equation for the Hamiltonian given in Eq.~(\ref{eq_H})
using explicitly correlated Gaussian (ECG) basis set expansion approach~\cite{blumermp}. 
After separating off the center-of-mass degrees of freedom, we expand the eigenstates of the
relative Hamiltonian in terms of ECG basis functions, which
depend on a number of nonlinear variational parameters that are
optimized through energy minimization~\cite{blumermp, blume09, blume11, yin15}.

{\em Bound state energy:} 
Trapped $(1,1)$ system, consisting one spin-up and
one spin-down particle has been solved analytically. 
The exact energy spectrum for zero-range interaction is given in Ref.~\cite{busch98}
while the contribution from the effective range is discussed in Ref.~\cite{werner12}. 
When $a_s$ is positive and much smaller than $a_\text{ho}$, the binding energy of the
two-particle pair with short-range interaction can be approximated by $E_s=-\hbar^2/(m a_s^2)$.
For $a_s=0.2a_\text{ho}$, the binding energy is approximately $-25E_\text{ho}$ and decreases with 
increasing $r_\text{eff}$.

For trapped two-component Fermi gases with more particles, i.e., $(N,N)$ systems, 
two particles with opposite spin form a molecule just like in $(1,1)$ system.
Such molecules were often treated as composite bosons
and an effective model with effective dimer-dimer interactions are shown to be
very accurate in the zero-range limit~\cite{petrov04, blume09, blume11}. 
In this effective model, the dimer-dimer interaction has a scattering length $a_{dd}=0.608a_s$,
a small positive value compared to $a_\text{ho}$, and does not support any bound state of two dimers~\cite{petrov04}.
This indicates that a dilute two-component Fermi gas
behaves like a Bose gas with hard-core repulsion in the BEC and zero-range limit.
However, dimer-dimer interaction is not $a$ $priori$ repulsive and cluster states can in principle exist.
An important question is what is the condition
for clusters to form.
In the following, we will show that the effective range $r_\text{eff}$ plays an important role.
 
\begin{figure}[htbp]
\includegraphics[angle=0,width=85mm]{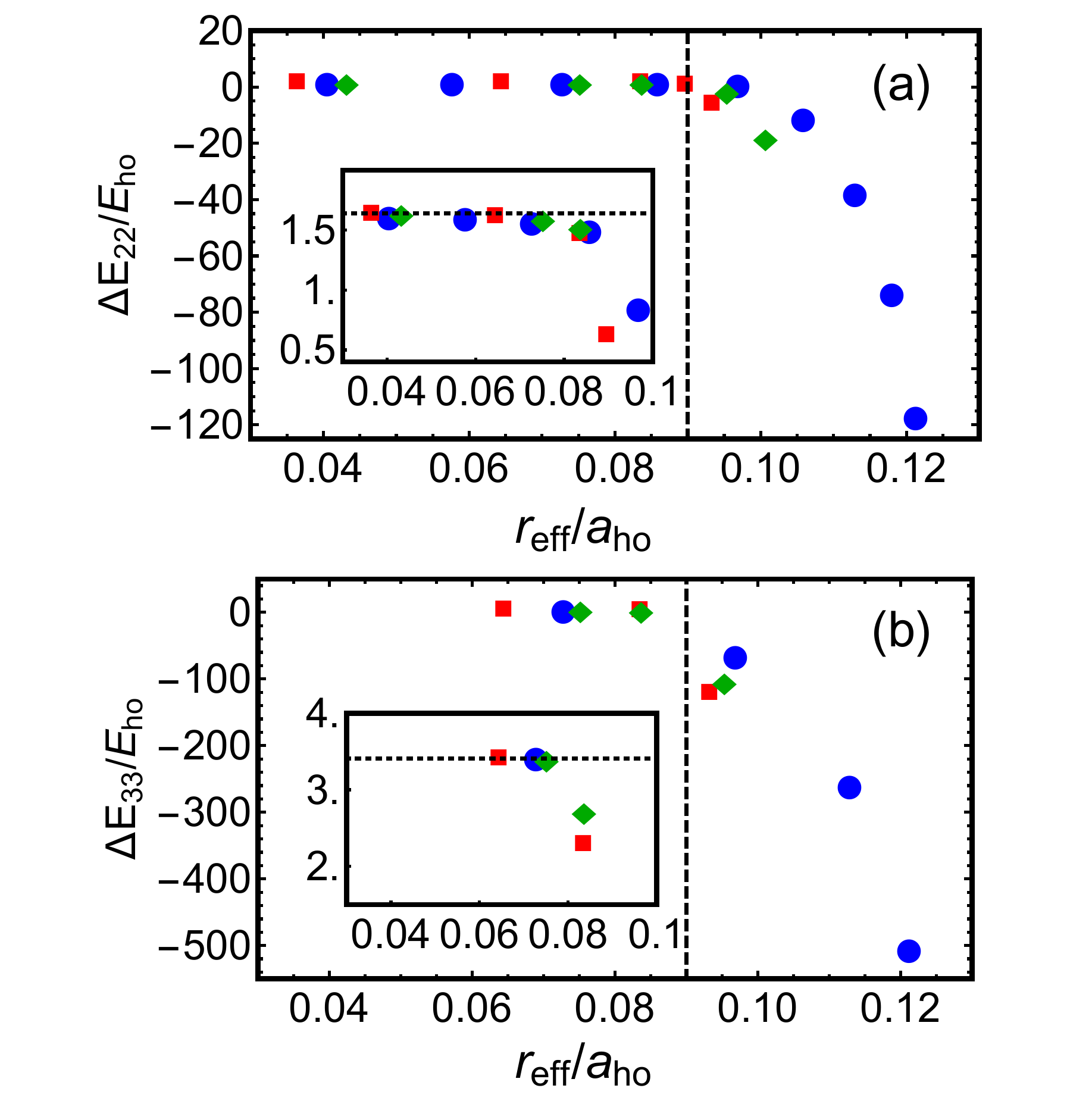}
\caption{(Color online) Panel (a) and (b) show the relative ground state energy for $(2,2)$ and $(3,3)$ systems
subtracting the dimer energies, $\Delta E_{22}=E_{22}-2E_{11}$ and $\Delta E_{33}=E_{33}-3E_{11}$, as a function
of effective range $r_\text{eff}$ for $a_s=0.2a_\text{ho}$. 
Blue circles, red squares, and green diamonds are calculated from potential (i), (ii), and (iii),
respectively. Dashed vertical line $r_\text{eff}=0.09a_\text{ho}$ marks the value of $r_\text{eff}$ where clusters start to form.
Insets show a magnified region with positive energy. Horizontal dotted line marks the energies calculated from effective
dimer model. 
}
\label{fig_energy}
\end{figure}

We calculate the relative ground state energies of $(1,1)$, $(2,2)$, and $(3,3)$ systems, $E_{11}$, $E_{22}$, and $E_{33}$, 
respectively, for fixed $a_s=0.2a_\text{ho}$ and a wide range of $r_\text{eff}$. The center-of-mass energy
is simply $3\hbar\omega/2$ for all systems and is not included in our results.
To identify the cluster formation between dimers, we follow Ref.~\cite{blume09} and subtract the energy of molecules from the four- and
six-body systems, i.e., we plot
$\Delta E_{22}=E_{22}-2E_{11}$ and $\Delta E_{33}=E_{33}-3E_{11}$ as a function of
$r_\text{eff}$ in Fig.~\ref{fig_energy}. Circles, squares, and diamonds are for potential (i), (ii), and (iii), respectively.
For small $r_\text{eff}$, our results agree with previous studies:
$\Delta E_{22}$ and $\Delta E_{33}$ are positive, and agree with the zero-range ground state energy for two and three weakly repulsive bosons 
with dimer-dimer scattering length $a_{dd}=0.608a_s$, which are marked as dotted lines in the insets.
This indicates that the ground state is indeed a gas-like state.
Energies $\Delta E_{22}$ and $\Delta E_{33}$ remain largely unchanged as $r_\text{eff}$ increases. 
As $r_\text{eff}$ approaches $0.09a_\text{ho}$, $\Delta E_{22}$ and $\Delta E_{33}$ decrease suddenly 
and turn negative for all three types of interactions, signalling the formation
of bound states between dimers. 
Note that the magnitude of $\Delta E_{33}$ is several times larger than $\Delta E_{22}$, indicating that the bound state in $(3,3)$ system
is indeed a three-dimer bound state, instead of a two-dimer bound state plus a single dimer.
Our calculations show that $r_\text{eff}$ at which cluster starts
to form is largely independent of the details of the two-body interactions and is roughly the same for (2,2) and (3,3) systems.
This shows that $r_\text{eff}$ is the deciding factor for cluster formation in two-component Fermi gases.

Our physics picture is as follows. The existence of cluster states in two-component Fermi gases
is determined by the counterbalance between the Fermi pressure between like-particles and the dimer-dimer interaction.
The Fermi pressure is a short-range effect, preventing like fermions from getting close to each other through
the requirement of anti-symmetrization. 
On the other hand, cluster formation requires all particles, like or unlike, to be at distances of the 
order of $r_\text{eff}$~\cite{petrov04}.
When $r_\text{eff}$ is small,
the anti-symmetrization of two like fermions cannot happen within such small distance.
Therefore, such interaction cannot support bound states between dimers.
When $r_\text{eff}$ becomes larger, the dimer-dimer interaction can potentially overcome the Fermi pressure,
allowing anti-symmetrization of two like fermions to happen within the range of $r_\text{eff}$.

\begin{figure}[htbp]
\includegraphics[angle=0,width=90mm]{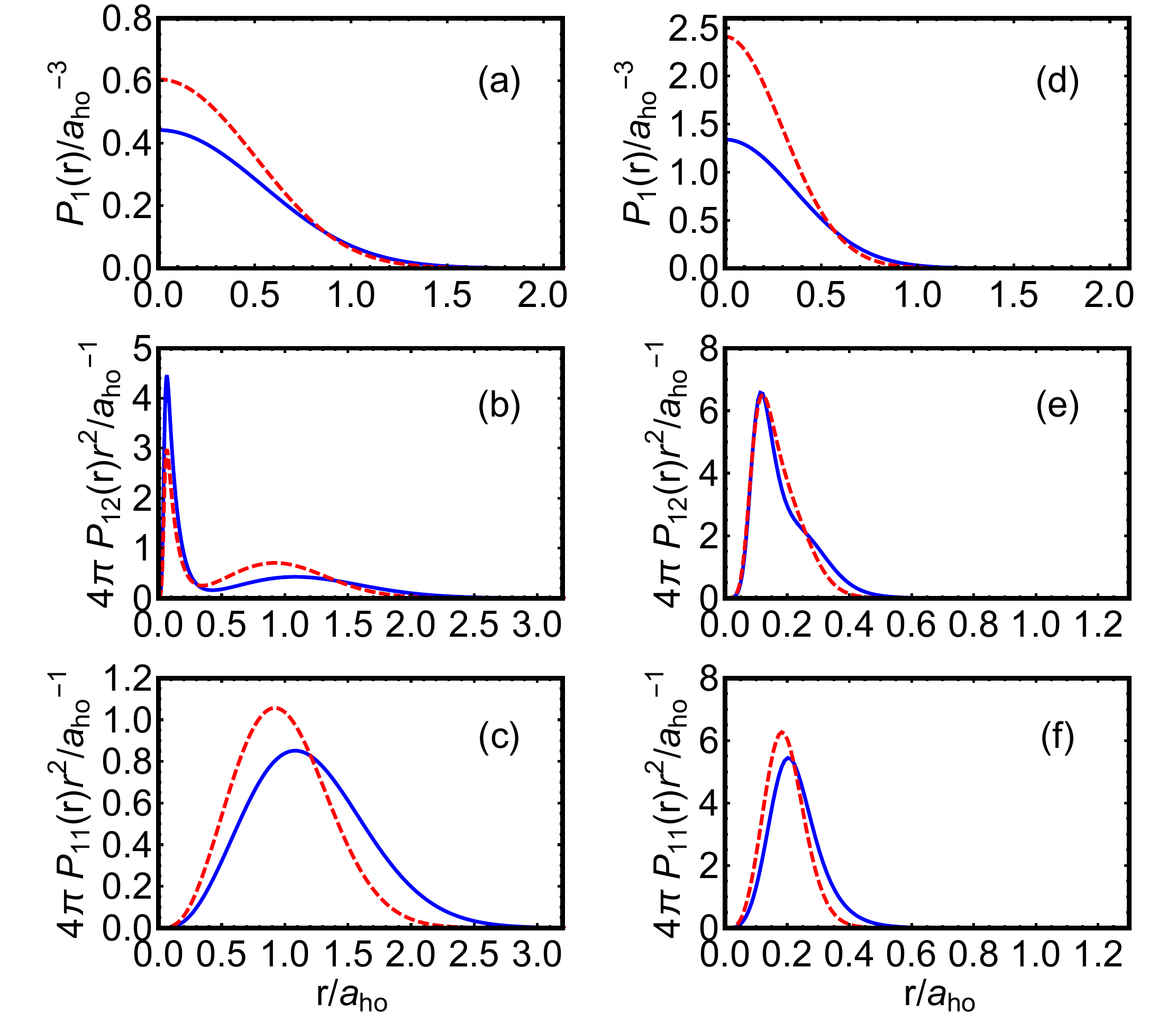}
\caption{(Color online) Solid and dashed lines show the structural properties for $(2,2)$ and $(3,3)$ systems,
respectively. (a), (b), and (c) show the radial density $P_1(r)$, scaled distribution function between unlike pair $4\pi P_{12}(r) r^2$,
and like pair $4\pi P_{11}(r) r^2$ for 
gas-like state with $a_s=0.2a_\text{ho}$ and $r_\text{eff}=0.073a_\text{ho}$.
(e), (f), and (g) show the same quantities as in (a), (b), and (c), but for 
cluster state with $a_s=0.2a_\text{ho}$ and $r_\text{eff}=0.113a_\text{ho}$.
}
\label{fig_sturcture}
\end{figure}

{\em Structural properties:} 
In order to take a peak at the wave function, we calculate several structural properties for both gas-like
and cluster states interacting through potential (i). 
First, we consider the spherically symmetric radial density $P_1(r)$, which tells the likelihood
of finding a particle at distance $r$ from trap center, with normalization $4\pi \int_0^{\infty} P_1(r) r^2 dr=1$.
Figure~\ref{fig_sturcture}(a) and (d) show the radial density $P_1(r)$
for gas-like and cluster state, respectively. For both states,
$P_1(r)$ peaks at the trap center and decays towards the edge. The overall extent of a single particle
can be measured by expectation value $\left\langle r \right\rangle=4\pi \int_0^{\infty} P_1(r) r^3 dr$. 
For both $(2,2)$ and $(3,3)$ systems,
the expectation values for the cluster states are about $30\%$ lower than gas-like states (see
Supplemental Material~\cite{supp}).
We confirm that the expectation values do not differ much for different $r_\text{eff}$ within the gas-like state
and cluster state. The drop in $\left\langle r \right\rangle$ represents a sudden change from gas-like
to cluster state.

Second, we consider the scaled distribution function between unlike pair $4\pi P_{12}(r) r^2$, which tells the
likelihood of finding two unlike particles at distance $r$ from each other, with normalization
$4\pi \int_0^{\infty} P_{12}(r) r^2 dr=1$. Since we consider the BEC limit, it is natural to expect
a sharp peak at a short distance that is of the order of $a_s$ as 
two unlike particles form a molecule. Indeed, for both the gas-like and
cluster states, a sharp peak at short distance is identified. A difference is that the distribution function for cluster state
vanishes at the order of trap length $a_\text{ho}$ while exhibits a lower and wider second peak for the gas-like state,
which corresponds to the distribution between two unlike particles that belongs to different dimers.

Third, we consider the scaled distribution function between like pair $4\pi P_{11}(r) r^2$, which tells the
likelihood of finding two like particles at distance $r$ from each other with normalization
$4\pi \int_0^{\infty} P_{11}(r) r^2 dr=1$. The distribution function between like pair is a good
gauge of the overall extent of the system. For gas-like state, a broad distribution centred around the
trap length $a_\text{ho}$ is observed, confirming that the system is indeed extended to the confinement of the trap.
For cluster state, we find that a peak appears at a similar distance to the unlike pair, which is an order
of magnitude smaller than gas-like state. This difference clearly
distinguishes the two states. Further details on the expectation values of distances between unlike and like pairs
are discussed in Supplemental Material~\cite{supp}.

\begin{figure}[htbp]
\includegraphics[angle=0,width=85mm]{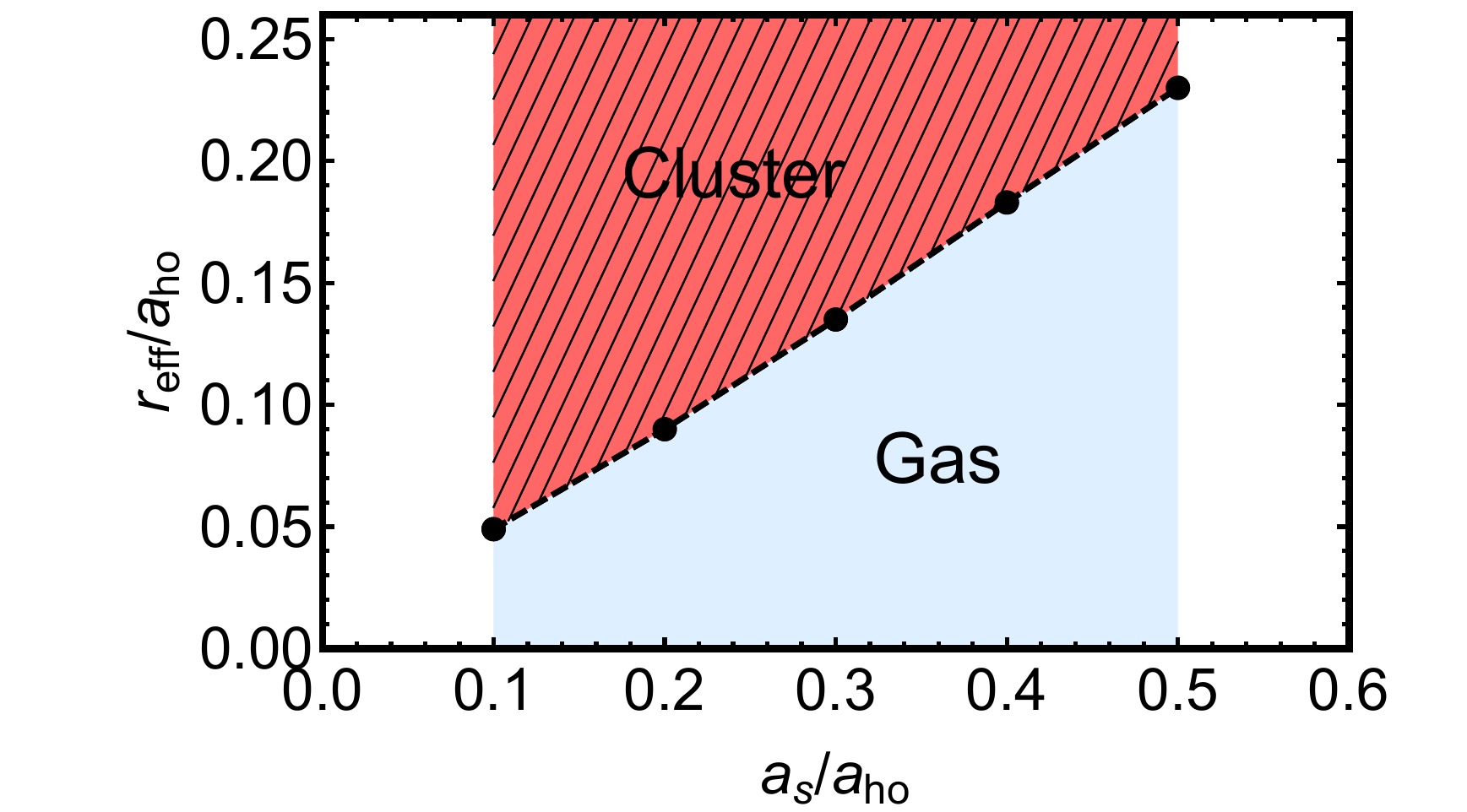}
\caption{(Color online) Boundary between the gas-like and cluster state
in the parameter space of scattering length $a_s$ and effective range $r_\text{eff}$. Hatched and solid
regions correspond to the cluster state and gas-like state, respectively.
}
\label{fig_phase}
\end{figure}

{\em Condition for cluster formation:}
For a certain $a_s$, both the ground state energies and the structural properties 
exhibit a clear transition between the gas-like and cluster state that is driven by the effective range $r_\text{eff}$. 
For each $a_s$, boundary is determined through the
sudden drop of $\Delta E_{22}$ [see, e.g., the rightmost red square in the inset of Fig.~\ref{fig_energy}(a)].
We then analyze $\Delta E_{33}$ and the expectation values from structural properties~\cite{supp} and find no discrepancy.
This allows us to map out the condition for cluster formation in Fig.~\ref{fig_phase}.

We make three observations.
First, the boundary between the two classes can be approximated by a linear relation $r_\text{eff}=0.46a_s$.
A shorter range interaction produces a hard-core repulsive dimer-dimer interaction
and results in a gas-like state. A longer range interaction, in contrast, can overcome the Fermi pressure and
leads to cluster formation.
Second, although we are not able to perform calculations for smaller $a_s$ or $r_\text{eff}$ due to numerical limitation,
if we assume that the linear relation continues to smaller $a_s$, a vertical line $r_\text{eff}=0$
will fall entirely into the gas-like state. This suggests that cluster states are absent for the zero-range interaction
and confirms the conclusion in Ref.~\cite{giogini04, blume07, blume09, blume11}. 
Third, as we increase $a_s$, the region of cluster state shrinks. This indicates that it could be more difficult to find
cluster states as the system moves towards the unitarity limit, i.e., with infinitely large scattering length.
This could explain the observation in Ref.~\cite{forbes12} where no cluster state was found for a wide range of $r_\text{eff}$.

Although we considered trapped few-body systems in this work, we expect many of our
findings to apply to larger and homogeneous systems (see
Supplemental Material~\cite{supp}). We also note that the transition between the gas-like and cluster state
happens within very small parameter range, and the changes in ground state energy and structural properties are drastic.
In the many-body limit, the transition could possibly
correspond to a first-order phase transition to a phase-separation phase~\cite{Krauth, Berciu}.

{\em Implications for experiments:}
Two-component Fermi gases in the BEC limit have been prepared in different cold atom systems. For example, 
a mixture of two hyperfine states, $|f, m_f \rangle=|9/2, -9/2\rangle$ and $|9/2, -5/2\rangle$, of $^{40}\text{K}$ have been
prepared across the BEC-BCS crossover~\cite{regal03}.
Other hyperfine states of $^{40}\text{K}$ ~\cite{regal04} and other species such as $^{6}\text{Li}$~\cite{julienne05} have also
been studied.

We consider 
neutral atom interactions modelled by van der Waals potential $V_{2b,\text{vdW}}(r_{ij})=-C_6/r_{ij}^6$. 
The relation between $r_\text{eff}$
and $a_s$ was found to be
$r_\text{eff}=\Gamma(1/4)^4/(6\pi^2) \bar{a} \left[1-2\bar{a}/a_s+2\left( \bar{a}/a_s\right)^2\right]$
~\cite{gao98} and shown to work reasonably well near broad Feshbach resonances~\cite{julienne14}.
Here, $\bar{a}=[4\pi/\Gamma(1/4)^2]R_\text{vdW}$, $R_\text{vdW}=(mC_6/\hbar^2)^{1/4}/2$,
and $\Gamma(x)$ is the Gamma function.
In the case of $^{40}\text{K}$, $R_\text{vdW}$ was calculated to be $64.90a_0$, where $a_0$ denotes the Bohr radius~\cite{kvdW}.
The scattering length is tunable near a Feshbach resonance between  
$|f, m_f \rangle=|9/2, -9/2\rangle$ and $|9/2, -5/2\rangle$, which is centred at $B_\text{pk}=224.21G$~\cite{regal03}.
We therefore determine that the system is in gas-like state for 
magnetic field $200.3G<B<233.9G$ (see
Supplemental Material~\cite{supp}).
Within this range, the two-component Fermi gas went through the BEC-BCS crossover
as observed in the experiments~\cite{regal03}.
We expect cluster to form outside this range.
This is consistent with the observations that two-component Fermi gases are generally stable near 
Feshbach resonances~\cite{petrov04}.

Cluster formation can potentially be observed through rf spectroscopy in a magnetically trapped two-component
Fermi gas with thousands of atoms~\cite{regal03, regal04}. Previous experimental measurements~\cite{regal03} on 
$^{40}\text{K}$ are performed with
$215G<B<230G$,
which lies entirely within the gas-like state according to our prediction. Our results suggest that cluster formation may be observed
further away from the Feshbach resonance.
A more direct reproduction of the system we studied, i.e., a trapped system with only few atoms,
can potentially be realized in optical tweezers~\cite{regal12, ni18} and optical microtraps~\cite{jochim11}.

{\em Final remarks:} 
We studied the formation of cluster state in two-component Fermi gases by numerically calculating the ground state energy
of a trapped few-body system interacting through various types of short-range potentials. Our findings corroborate the
picture that the counterbalance between the Fermi pressure and the short-range interactions determines the boundary between
the gas-like and cluster state.
Although our calculations are performed with short-range interactions due to numerical restrictions, we expect such 
result to apply to realistic interactions because of the short-range nature of Fermi pressure.
We also provide an estimate for the parameter regime where such cluster states could potentially be identified in the experiments.

\begin{acknowledgements}
We are grateful for discussions with Jia Wang and Yangqian Yan.
This research was supported by the Australian Research Council (ARC) Discovery Programs, 
Grants No. DP170104008 (H.H.), No. FT140100003 and No. DP180102018 (X.-J.L).
\end{acknowledgements}

\bibliographystyle{apsrev4-1}
\bibliography{mybib}

\end{document}


\title{Supplemental Material for ``Cluster formation in two-component Fermi gases''}
\author{X. Y. Yin}
\affiliation{Centre for Quantum and Optical Science, Swinburne University of Technology,
Melbourne, Victoria 3122, Australia}
\author{Hui Hu}
\affiliation{Centre for Quantum and Optical Science, Swinburne University of Technology,
Melbourne, Victoria 3122, Australia}
\author{Xia-Ji Liu}
\affiliation{Centre for Quantum and Optical Science, Swinburne University of Technology,
Melbourne, Victoria 3122, Australia}

\maketitle

\section{Deeper bound states supported by model potentials}
In this work, we only consider attractive potentials (i), (ii), and (iii) that support one $s$-wave bound state in free space.
Here, we explain the reasoning for this consideration and discuss the relevance of deeper bound states supported 
by these potentials.

When these attractive potentials support only one $s$-wave bound state, the bound state is fairly universal, i.e., the binding
energy depends mostly on $a_s$ and $r_\text{eff}$. For example, the relative energies for trapped $(1,1)$ system with 
$a_s=0.2a_\text{ho}$ and $r_\text{eff}=0.06a_\text{ho}$ interacting with potentials (i), (ii), and (iii) are $-37.24E_\text{ho}$,
$-37.45E_\text{ho}$, and $-37.35E_\text{ho}$, respectively, which differ by at most $0.6\%$.

When these potentials becomes deeper, they support successively more two-body bound states. These deeper bound states have much
lower energies and their energies depend highly on the details of the potentials other than $a_s$ and $r_\text{eff}$. 
In other words, the properties of these deeper bound states are highly non-universal. We consider again 
trapped $(1,1)$ system with 
$a_s=0.2a_\text{ho}$ and $r_\text{eff}=0.06a_\text{ho}$ interacting with potentials (i), (ii), and (iii) that support exactly two
$s$-wave bound states in free space. The relative energies for the ground states are $-24589.2E_\text{ho}$, $-8495.09E_\text{ho}$, 
and $-8891.35E_\text{ho}$, respectively. As we can see, the ground state energies are vastly different, especially for
potential (i). This is expected since the repulsive barrier can have a large impact on non-universal properties.

We also note that when the potentials support two bound states, the absolute value of binding energy of the ground state is $200-600$ times
higher than the first excited state. If realistic interactions in the experiments support more than one bound state, we expect them 
to be well separated in the relevant parameter regime. Hence, the deeper bound states should not play a significant role on the
experimentally relevant time scale.

Since the deeper bound states are highly non-universal, the properties of two-component Fermi gases with more particles cannot 
be described by the framework in Ref.~\cite{petrov04}. Naturally, our condition for cluster formation, which depends only on $a_s$
and $r_\text{eff}$, also do not apply to these states.

\section{Chemical potential for cluster states}

\begin{figure}[htbp]
\includegraphics[angle=0,width=80mm]{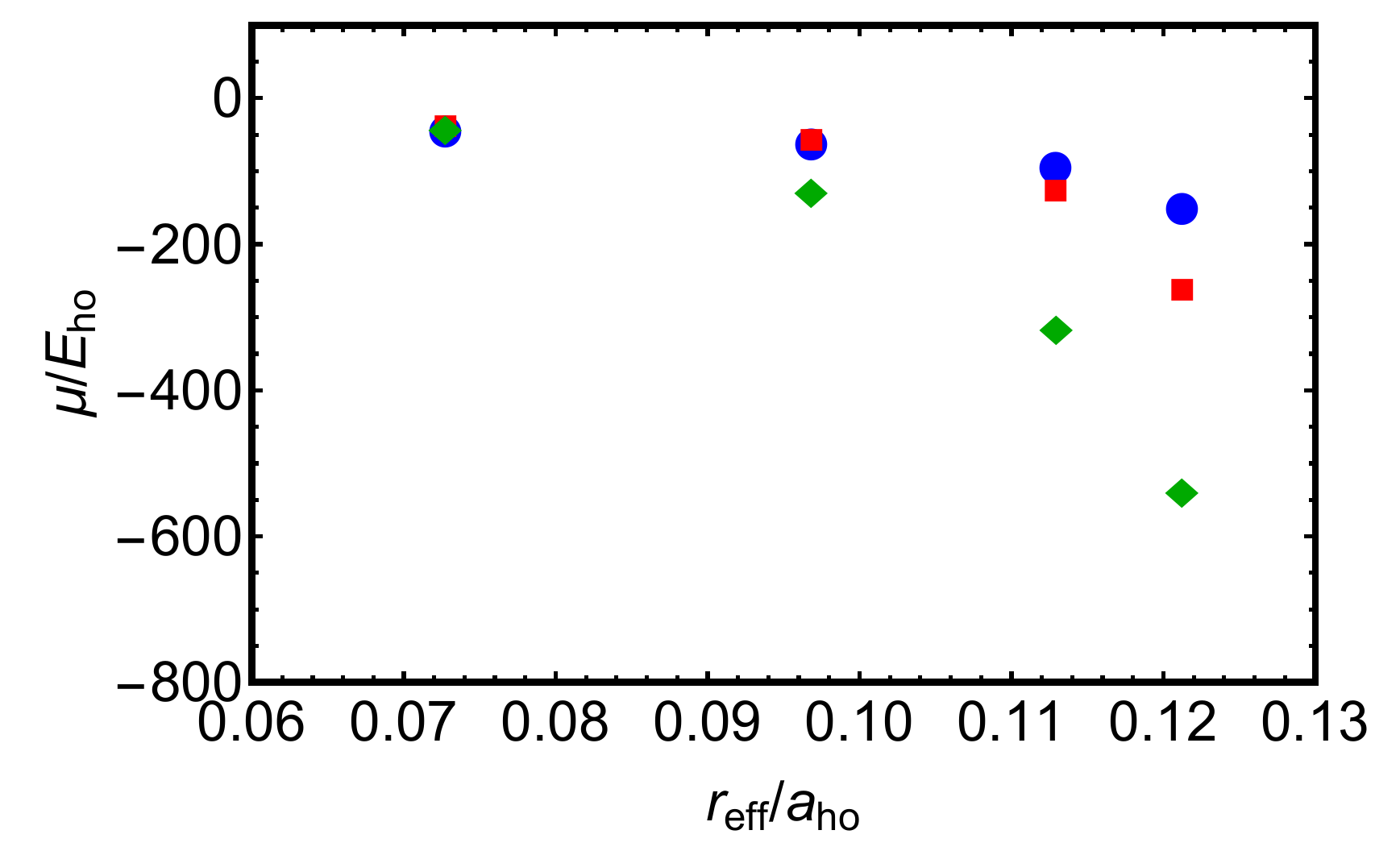}
\caption{(Color online) Blue circles, red squares, and green diamonds show the
relative ground state energy of $(1,1)$ system $E_{11}$, chemical potential of $(1,1)$ system $\mu_{11}$,
and chemical potential of $(2,2)$ system $\mu_{22}$, respectively. Calculations are done for type (i) potential
with $a_s=0.2a_\text{ho}$.
}
\label{fig_mu}
\end{figure}

In this work, we consider $(1,1)$, $(2,2)$, and $(3,3)$ systems. To provide further insights into larger systems,
we calculate the chemical potential. The chemical potential for few-body systems is
defined as the energy change when adding one particle. For two-component Fermi gases, we consider
$\mu=\mu_\uparrow+\mu_\downarrow$, that is the energy change when adding a two-particle pair.
Specifically, we show $\mu_{11}=E_{22}-E_{11}$ and $\mu_{22}=E_{33}-E_{22}$ in Fig.~\ref{fig_mu}. We additionally 
plotted $E_{11}$, which corresponds to the energy change when adding a pair to the vacuum.

For gas-like state, the chemical potential remains a constant when number of particles increases. This constant
is approximately equal to the energy of a dimer.
For cluster state, we find that the absolute value of chemical potential increases with number of particles for the 
few-body systems that we considered. However, it is difficult to predict whether the chemical potential will 
eventually saturate to a constant based solely on few-body results.

\section{Applicability to the homogeneous system}

\begin{figure}[htbp]
\includegraphics[angle=0,width=80mm]{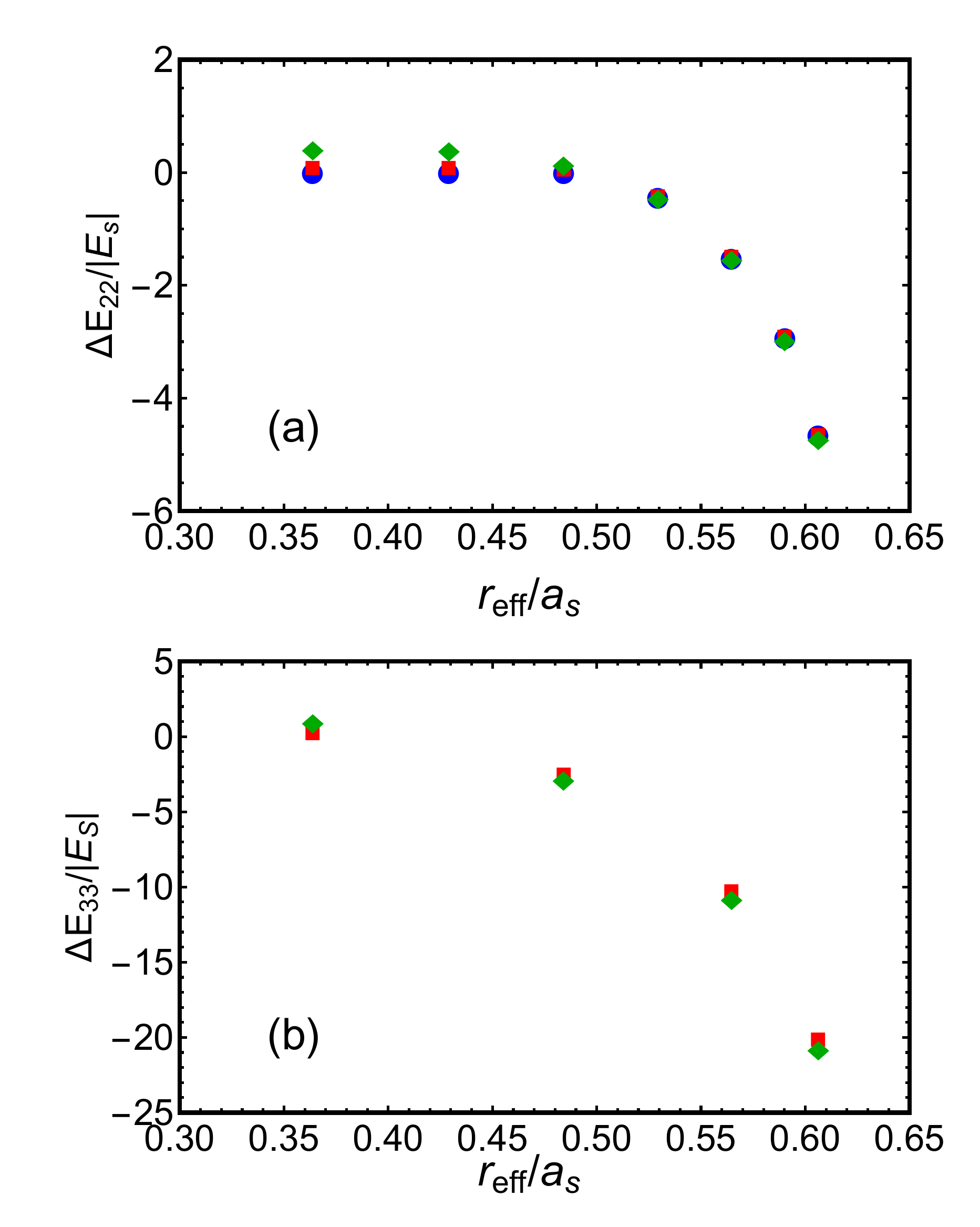}
\caption{(Color online) Panel (a) and (b) show the relative ground state energy for $(2,2)$ and $(3,3)$ systems
subtracting the dimer energies, $\Delta E_{22}=E_{22}-2E_{11}$ and $\Delta E_{33}=E_{33}-3E_{11}$, as a function
of effective range $r_\text{eff}$ for type (i) potential.
Lengths and energies are expressed in the unit of $a_s$ and $|E_\text{s}|=\hbar^2/(m a_s^2)$,
respectively.
Blue circles [$(2,2)$ only], red squares, and green diamonds are 
for $a_s=0.1a_\text{ho}$, $0.2a_\text{ho}$,
and $0.5a_\text{ho}$, respectively.
}
\label{fig_scale}
\end{figure}
 
The calculations in this work are done for a trapped system.
Our main results are reported in the harmonic trap length 
$a_\text{ho}$ and the harmonic oscillator energy $E_\text{ho}$, which are the
natural units in trapped systems.
An important question is that, how does our results in a trap connect to
a homogeneous system. 

For gas-like state, the average distance between like particles are of the order of $a_\text{ho}$. 
The energies are determined, to the leading order, by the trapping
potential. The trap length is the most important length scale in this case.

For cluster state, the average distance between like particles are much smaller than
$a_\text{ho}$. Because the size of the cluster state is an order of magnitude smaller than
the $a_\text{ho}$, it is natural to expect that $a_\text{ho}$ becomes an
irrelevant length scale. In this case, the most relevant length scale is set by the $s$-wave
scattering length $a_s$ and the energy scale by the 
absolute value of zero-range binding energy $|E_\text{s}|=\hbar^2/(m a_s^2)$.

To examine if $a_\text{ho}$ truly becomes irrelevant, we plot the relative ground
state energies $\Delta E_{22}$ and $\Delta E_{33}$ calculated with different
$a_s/a_\text{ho}$ in Fig.~\ref{fig_scale} in terms of scattering unit $a_s$ and $|E_\text{s}|$.
We find the energies for cluster state roughly collapse. This indicates that
for the parameter regime that we considered, the impact of the trapping potential
to the energy of cluster state is small. Therefore, we expect that our condition
for cluster formation and the binding energy for cluster state should also apply to the
homogeneous system.

\section{Expectation values related to structural properties}
To gain more insights from the structural properties, we consider three expectation
values calculated from the spherically symmetric radial density $P_1(r)$, 
the scaled distribution function between unlike pair $4\pi P_{12}(r) r^2$,
and the scaled distribution function between like pair $4\pi P_{11}(r) r^2$.
The expectation value $\left\langle r \right\rangle=4\pi \int_0^{\infty} P_1(r) r^3 dr$ measures the overall extent of a single particle,
while $\left\langle r_{12} \right\rangle=4\pi \int_0^{\infty} P_{12}(r) r^3 dr$ and 
$\left\langle r_{11} \right\rangle=4\pi \int_0^{\infty} P_{11}(r) r^3 dr$ measure the expected distance between
unlike and like particles, respectively.

\begin{figure}[htbp]
\includegraphics[angle=0,width=80mm]{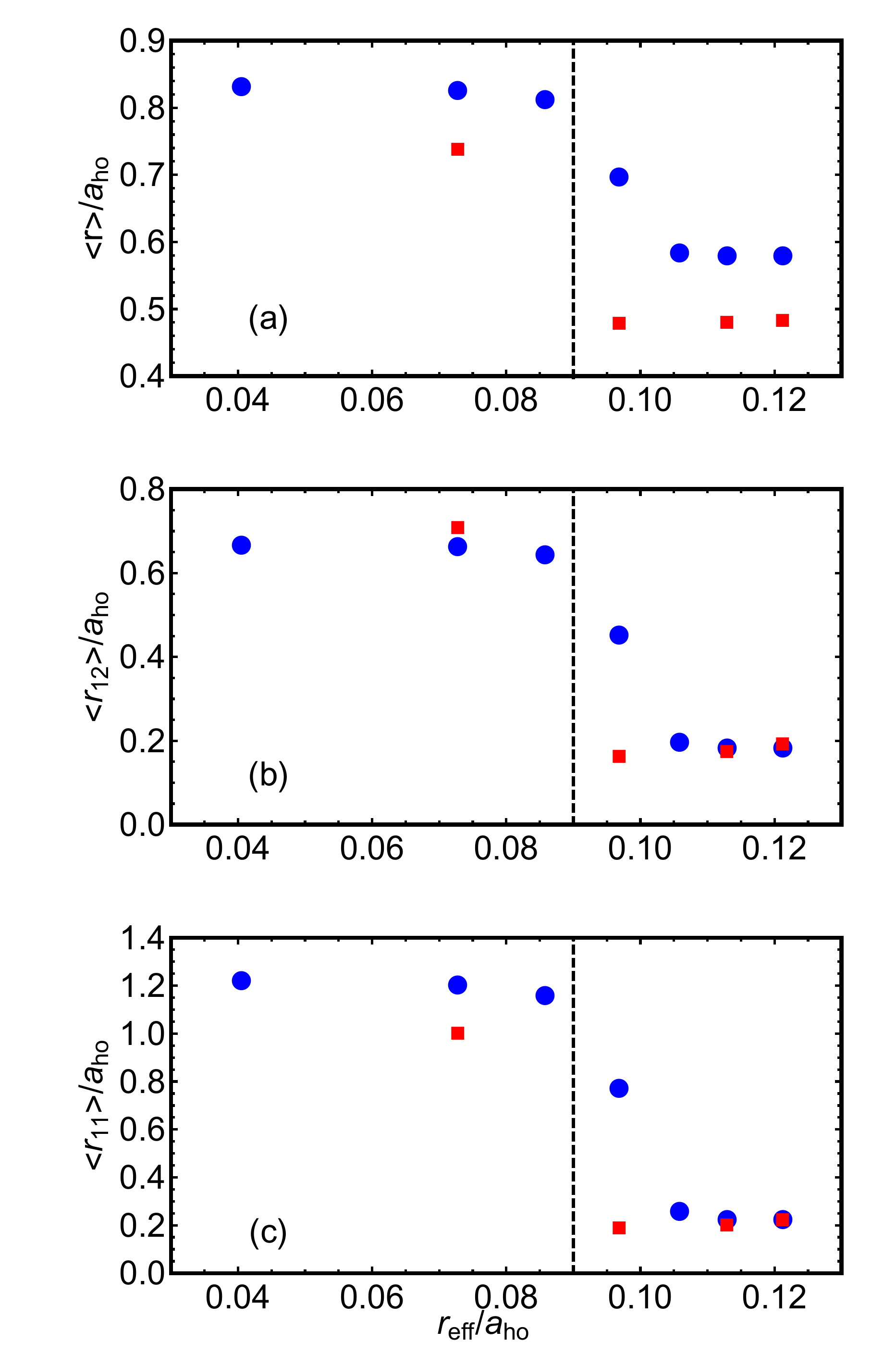}
\caption{(Color online) Panel (a), (b), and (c) show the expectation values $\left\langle r \right\rangle$, $\left\langle r_{12} \right\rangle$,
and $\left\langle r_{11} \right\rangle$ as a function of effective range $r_\text{eff}$ for $a_s=0.2a_\text{ho}$. Blue circles and
red squares are for $(2,2)$ and $(3,3)$ systems, respectively. Dashed vertical line $r_\text{eff}=0.09a_\text{ho}$ 
marks the value of $r_\text{eff}$ where clusters start to form.
}
\label{fig_supp}
\end{figure} 

Circles and squares in Fig.~\ref{fig_supp} show the three expectation values for $(2,2)$ and $(3,3)$ systems, respectively.
In both systems, a sudden decrease of all three expectation values occurs at around $r_\text{eff}=0.09a_\text{ho}$,
where the transition from gas-like state to cluster state occurs. 
This value is consistent with the the transition point determined through bound state energy in the main text.
Notably, $\left\langle r_{11} \right\rangle$ decreases by
an order of magnitude as cluster forms. 

\section{Additional details on the Feshbach resonance in ${}^{40}K$}
\begin{figure}[htbp]
\includegraphics[angle=0,width=75mm]{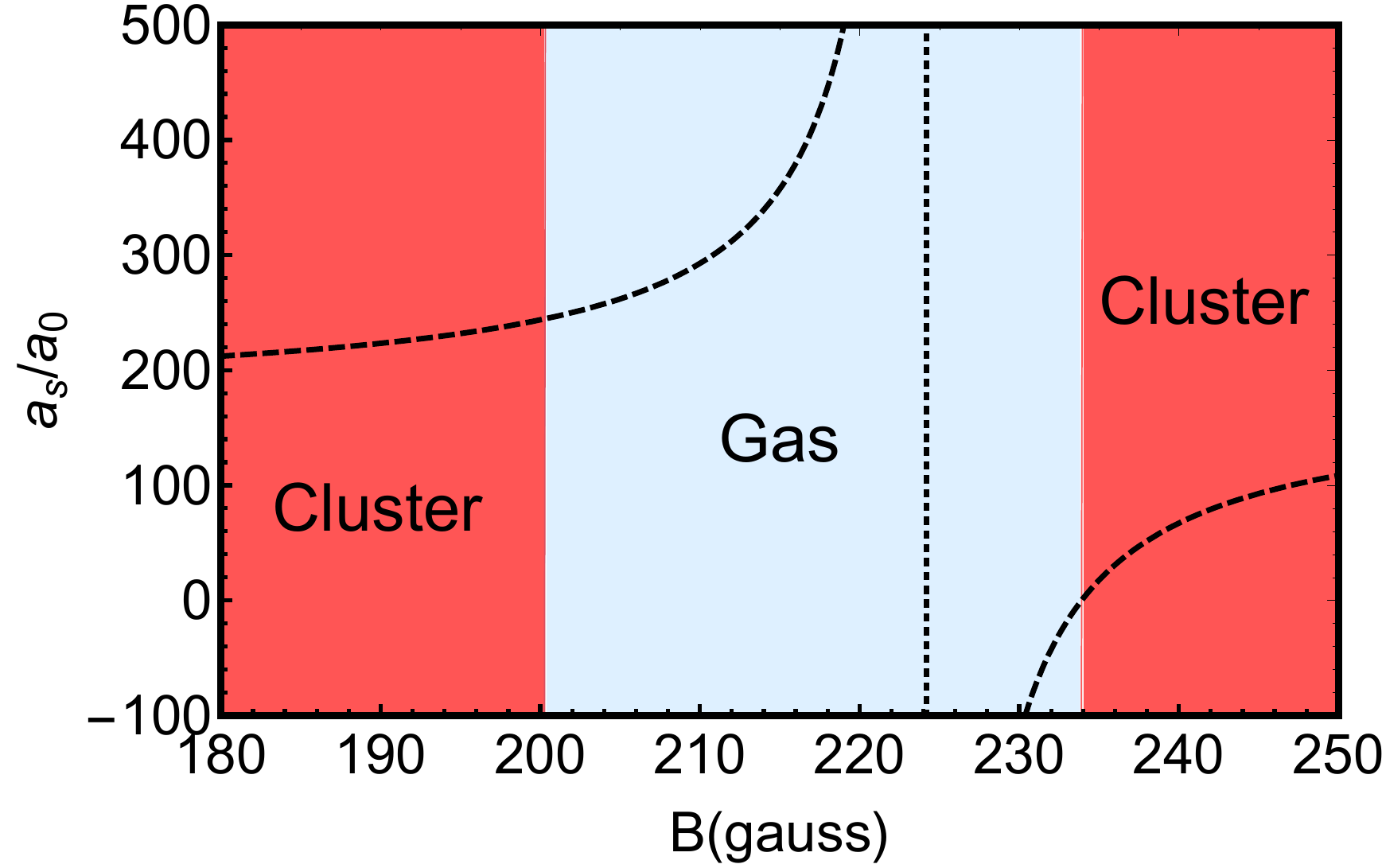}
\caption{(Color online) Dashed line shows $a_s$ as a function of magnetic field $B$ near the Feshbach resonance between
$|f, m_f \rangle=|9/2, -9/2\rangle$ and $|9/2, -5/2\rangle$ of $^{40}\text{K}$. Dotted vertical line marks the
position of Feshbach resonance. The regime of cluster state is determined according to condition $0<r_\text{eff}<0.46a_s$.
}
\label{fig_exp}
\end{figure}

We consider the Feshbach resonance between
$|f, m_f \rangle=|9/2, -9/2\rangle$ and $|9/2, -5/2\rangle$ in ${}^{40}K$~\cite{regal03}. 
We plot the relation between $a_s$ and the magnetic field $B$ in Fig.~\ref{fig_exp} according to the 
background scattering length $a_\text{bg}=174a_0$, resonance peak $B_\text{pk}=224.21\pm0.05G$
and width $w=9.7\pm0.6G$, measured with high precision in Ref.~\cite{regal03}.
Together with relation between $r_\text{eff}$ and $a_s$ discussed in the main text, 
we can determine the boundary between the cluster state and the gas-like state for this Feshbach resonance.

\bibliographystyle{apsrev4-1}
\bibliography{mybib}